\documentclass[12pt,prb,aps]{revtex4}

\usepackage{amsmath}
\usepackage{amsmath,amsfonts,amssymb,braket,epsfig}
\begin{document}

\title{Bloch and Josephson Oscillations in a Ring of an Ideal Bose Gas}
\author{Leon Gunther}
 \affiliation{Department of Physics and Astronomy\\ Tufts
University\\ Medford, Massachusetts 02155}

\begin{abstract}
  We show that an Ideal Bose gas that is contained within a very thin ring exhibits phenomena analogous to the Bloch and Josephson oscillations of a charged Ideal Fermi gas in a thin ring. If the walls of the ring are constrained to have an angular velocity $\omega$, the angular momentum has an anomalous component that is periodic in $\omega$, with a period equal to the quantum of angular velocity $\omega_{0}\equiv \hbar /mR^{2}$. If a constant applied torque is applied to the walls, there will be component of the angular momentum of the gas that is periodic in time, with a 'Josephson frequency' given by $f _{J} =\tau/N \hbar$. Finally, we show that the oscillations are an automatic feature of the quantum regime of any ring of an ensemble of identical particles, even with particle interactions.\\

\end{abstract}

\maketitle

\section{INTRODUCTION}
This paper is based upon the research that Joe and I carried out around 1970. Motivated by the papers of Byers and Yang \cite{by} and Bloch \cite{bloch}, we  demonstrated that the Bloch oscillations associated with a superconductor were present in a thin ring of an Ideal Fermi gas, albeit with a flux quantum determined by a single electron charge (thus, $hc/e$), in contrast with $hc/2e$ for the BCS superconductor.   We also showed that there were analogous oscillations in the behavior of a rotating ring of an ideal Bose gas. None of our results were published. However, the results for the Ideal Fermi gas ultimately were and still are being developed much further for a real metal in the normal state (with important contributions made by Joe and his coworkers), including elastic scattering of electrons off impurities, electron scattering off magnetic impurities, electron-electron scattering, and inelastic electron-phonon scattering.  See Joe's monograph \cite{joe1} for more details as well as the very recent paper by Joe and collaborators \cite{ham}.  The essential predictions for the normal metal ring were ultimately observed in a number of beautiful experiments \cite{levy}.  On the other hand, the work on the Bose gas was left in the bins of old notes. The purpose of this paper is to revive this problem and to honor my relationship with Joe.

One of the key steps in the formulation of this problem is, for me at least, based upon a talk by Gordon Baym presented at the 1967 Summer School held at St. Andrews, Scotland on "Mathematical Methods in Solid State and Superfluid Theory".\cite{andrews} Baym pointed out that if a fluid is placed in a rotating cylinder, the energies in the partition function must be those with respect to the rotating frame of reference.  We will later see how this requirement leads to Bloch and Josephson oscillations in a ring of an Ideal Bose gas.

By coincidence, it was at this Summer School that I was fortunate to meet Joe. We had a marriage of like minds immediately. Joe was on his way to the US for a leave of absence, fresh from his work on demonstrating that while a one-dimensional Ising model with short range interactions does not exhibit a phase transition in the thermodynamic limit (that is, as the number of spins N goes to infinity), a finite chain of spins could exhibit long range order that is not significantly weaker than that obtained in the thermodynamic limit in a system with long range interactions . I had come to other corresponding conclusions about finite systems during my post-doc at Orsay that year. There I had heard Bernard Jancovi\c{c}i \cite{jan} give a talk on novel behavior of the susceptibility of a finite 2D harmonic lattice in spite of the accepted result that a 2D harmonic lattice was unstable. His analysis revealed that the long wavelength divergence of fluctuations that destroyed long range order were cut off in a finite system. That same year, I had also heard Vladimir Tkachenko \cite{tka} give a talk about his results demonstrating that a rotating lattice of vortices in a superfluid was unstable due to the diverging fluctuations in the vibrations of the vortices. I was led to note that the divergence vanished in a finite lattice, due to a cutoff in the long-wavelength fluctuations that was proportional to log-N.

 Our meeting at St. Andrews led to many years of  wonderful times together as friends as well as in research that reflected our shared view that theory must always strive to honestly reflect experimental conditions.  I believe that this philosophy was the foundation for Joe's incredible ability to translate theory into a language that experimentalists could not only understand but also use to produce wonderful experimental results.  Joe would often come up with great suggestions for new areas of research. However, for me it was even more delightful that whenever I had an idea, Joe was there to analyze it and lead us with an explosion of further developments, ways of understanding the phenomena and simplifying the analysis.

I had the pleasure of working with Joe and his colleagues on research that revealed that the study of finite systems could not only make some non-existence theorems based upon infinite systems of not too great importance practically, but also teach us much about systems in the thermodynamic limit and reveal novel behavior in real, finite systems.
We take all this for granted since the rise of nanoscopic physics in the '70s.

\section{Comparison of a Ring of an Ideal Fermi Gas with a Ring of an Ideal Bose Gas}

\subsection{Energy of the Quantum States}
Our ring has a radius R and cross-sectional dimensions that are so small that the first excited state with respect to either dimension has an energy much larger than the thermal energy kT. As a result, the ring is effectively one-dimensional. The rotational momentum is quantized in the inertial lab frame. Thus we have

\begin{equation}
p=n\frac{\hbar}{R}
\label{eq:mom}
\end{equation}
where n is any integer (including zero).

For both the Ideal Fermi gas, as well as the Ideal Bose gas (in the rotating frame of reference of a rotating ring), the energy of the states (simply kinetic energy) that is relevant in the partition function can be expressed as

\begin{equation}
E_{n}=\frac{p_{n}^2}{2m} =\frac{\hbar^{2}(n-\phi)^2}{2mR^2}
\label{eq:kin}
\end{equation}
where m is the particle mass and n is any integer from $-\infty$  to $+\infty$.
Note that for the Fermi case, this energy is in the lab frame, while for the Bose case, this energy is in the rotating frame.

In the case of the \textbf{Fermi} gas, the charged particles are in the presence of an external magnetic field parallel to the axis of the ring.  The parameter $\phi$ is the ratio

\begin{equation}
\phi=\frac{\Phi}{\Phi_{0}}
\end{equation}
where $\Phi$ is the total magnetic flux through the ring and $\Phi_0$ is the flux quantum given by

\begin{equation}
\Phi_0=\frac{hc}{e}
\end{equation}
The total flux is a sum of the flux due to the external field and the flux produced by an electric current of the charged particles in the ring.

In the case of the \textbf{Bose} gas, the situation is a bit more complex. We have a ring of gas that is bounded by a wall that is rotating at a fixed angular velocity $\omega$. The rotational momentum is quantized in the inertial lab frame, and is thus given by equation (\ref{eq:mom}).
There is a corresponding quantum of velocity $v_{0}$ and of angular velocity $\omega_{0}$

\begin{equation}
v_{0}=\frac{\hbar}{mR}  ~~~~~~~~~
 \omega_{0}=\frac{\hbar}{mR^{2}}
\end{equation}

We assume that the system has come to thermodynamic equilibrium. As we mentioned above, the energy in the partition function is then the energy in the rotating frame. The rotational velocity in the rotating frame is given by

\begin{equation}
v_{rot}=v_{lab}-\omega R
\end{equation}

Thus the unit of quantum velocity in the rotational frame is

\begin{eqnarray}
v_{n}&=&n \frac{\hbar}{mR} - \omega R \\
\end{eqnarray}

Thus we obtain for the Bose gas equation (\ref{eq:kin}), with
\begin{equation}
\phi=\frac{\omega}{\omega_{0}}
\end{equation}

Note that the energy of a state can be written as

\begin{equation}
E_{n}=\frac{\hbar \omega_{0}}{2} (n-\phi)^2
\label{eq:kin2}
\end{equation}

 Thus, $e_{0}\equiv \hbar \omega_{0}/2$ is the characteristic energy of the system.

\subsection{Electric Current in the Fermi Gas}
In the case of the Fermi gas, it is  the electric current that exhibits Bloch oscillations as a function of the flux. It is given by

\begin{eqnarray}
I&=&\frac{e}{2\pi R}\sum_{n=-
  \infty}^\infty v_{n}F_{n}\\
  &=&\frac{e}{2\pi R}\sum_{n=-\infty}^\infty \frac{\hbar}{mR}(n-\phi)F_{n}
=\frac{e\hbar}{2\pi mR^2}\sum_{n=-\infty}^\infty (n-\phi)F_{n}
\label{eq:current}
\end{eqnarray}
where $F_{n}$ is the Fermi function

\begin{equation}
F_{n}=\frac{1}{exp[(E_{n}-\mu)/kT] +1}
\end{equation}
Here $\mu$ is the chemical potential.

From equation (\ref{eq:current}) we see that the current is a periodic function of the magnetic flux, with a period equal to the flux quantum $\Phi_{0}=hc/e$.

\subsection{Angular Momentum in the Bose Gas}
For the Bose gas we are interested in how the angular momentum  $L$ depends upon the fixed angular velocity. In the lab frame it is given by

\begin{eqnarray}
L=\hbar\sum_{n=-\infty}^\infty n f_{n}
\end{eqnarray}
where $f_{n}$ is the Bose function

\begin{equation}
f_{n}=\frac{1}{exp[(E_{n}-\mu)/kT] -1}
\end{equation}
Here $\mu$ is the chemical potential.

 We can separate the total angular momentum into two parts:
\begin{eqnarray}
L=L_{class}+L_{anom}
\end{eqnarray}
The first part is what we obtain in the classical limit:

\begin{eqnarray}
L_{class}= \mathcal{I} \omega =N\hbar \phi = \hbar\sum_{n=-\infty}^\infty \phi f_{n}
\end{eqnarray}
where $\mathcal{I}=NmR^{2}$ is the total moment of inertia.

The second part is the anomalous part - the angular momentum in the rotating frame - which would normally vanish since then the walls would carry the entire gas along with it.

\begin{eqnarray}
L_{anom}=\hbar\sum_{n=-\infty}^\infty (n-\phi)f_{n}
\label{eq:angmom}
\end{eqnarray}

From equation (\ref{eq:angmom}) we see that the anomalous angular momentum is a periodic function of the applied angular velocity, with a period equal to the quantum of angular velocity, $\omega_{0}=\hbar/{mR^{2}}$.  Thus, the anomalous angular momentum corresponds to the above Bloch oscillations of the electric current.

\subsection{Crossover Temperatures for Oscillations}
It is clear that the discreteness of the energy levels is responsible for the observability of Bloch oscillations. As we raise the temperature, the discreteness becomes less relevant. Generally, the energy level spacing is given by
\begin{eqnarray}
\Delta  E_{n}= \left[(n+1)^2-n^2\right]e_{0}=(2n+1)e_{0}
\end{eqnarray}

In the \textbf{classical regime}, the average energy per particle is $\sim kT$. The crossover temperature is determined by setting the characteristic energy level spacing equal to $kT$. The corresponding quantum number is $n=\sqrt{kT/e_{0}}$. Thus, the relevant energy level spacing is $~2e_{0}\sqrt{kT/e_{0}}=2\sqrt{kT e_{0}}$. And finally, we obtain the crossover temperature $T_{C}$ from the equation $\sqrt{kT e_{0}}=kT$:
\begin{eqnarray}
T_{C}\sim e_{0}/k
\end{eqnarray}

For \textbf{Fermions}, the characteristic energy is the Fermi energy $\epsilon_{F}$, which corresponds to a quantum number $n\sim N/2$. Thus the relevant energy level spacing is $\Delta  E_{N/2} \sim Ne_{0}$. Again, the crossover temperature $T_{F}$ is determined by $\Delta  E=kT$, so that it is given by
\begin{eqnarray}
T_{F}\sim Ne_{0}/k
\end{eqnarray}
In fact, it can be shown that the persistent current is given by \cite{chempot}
\begin{eqnarray}
I=N \sum_{p=1}^\infty \frac{(-1)^{Np}}{2 \pi p}\frac{2\pi ^2 pkT/Ne_{0}}{\sinh[2\pi ^2 pkT/Ne_{0}]}
\end{eqnarray}
Hence, the actual crossover temperature is better represented by
\begin{eqnarray}
T_{F}= Ne_{0}/(2 \pi ^{2} k)
\end{eqnarray}

For \textbf{Bosons}, the situation is entirely different. We know that bosons are attracted into the same state. The 3D Bose gas undergoes a Bose-Einstein condensation. While there is no condensation in 1D or 2D, there is nevertheless a relatively high occupation of the states with low energies - with quantum number n of order unity - as a result of the minus sign in the denominator of the distribution function.

We can learn a lot from an analysis of the situation at essentially absolute zero. In this case, all particles are in the ground state.  Let $\alpha(\phi) \equiv -\mu (\phi)/kT$

Then we must have
\begin{equation}
f_{0}=\frac{1}{exp[e_{0} \phi^2/kT +\alpha] -1}=N
\end{equation}

For simplicity, we take $\phi=0$. Then,
\begin{equation}
\alpha =\ln(1+1/N) \sim 1/N
\end{equation}
For low temperatures, we still expect $\alpha \sim 1/N$.
We want $f_{n\neq0}$ to be of order unity for n on the order of unity but negligibly small for n not of order unity.

Since
\begin{equation}
f_{n}=\frac{1}{exp[e_{0} n^2/kT +\alpha] -1} \sim\frac{1}{exp[e_{0} n^2/kT + 1/N] -1}
\end{equation}
we must have $kT<Ne_{0}$. Thus, the crossover temperature $T_{B}$ for our Bose gas is the same as it is for the Fermi gas, namely
\begin{eqnarray}
T_{B}\sim Ne_{0}/k
\end{eqnarray}
It is more straightforward to show that for the low energy states to dominate, we must have 
\begin{eqnarray}
f_{n} = \frac{1}{exp[(e_{0}(n-\phi)^2-\mu)/kT] -1} \cong \frac{1}{(e_{0}(n-\phi)^2-\mu)/kT}
\label{eq:approx}
\end{eqnarray}
for the dominant states.\\

We will define a characteristic temperature
\begin{eqnarray}
T_{B} \equiv Ne_{0}/2 \pi ^{2}k
\end{eqnarray}

The chemical potential is determined by the N, T and $\phi$ through the equation

\begin{equation}
N=\sum_{n=-\infty}^\infty f_{n} \cong \sum_{n=-\infty}^\infty   \frac{1}{(e_{0}(n-\phi)^2-\mu)/kT}
\end{equation}

We will use the approximate expression for N from now on. Let us introduce the reduced \textbf{ reduced chemical potential}  $m \equiv \mu /e_{0}$ and the \textbf{reduced temperature} $t \equiv T/ T_{B} $. Then it is straightforward to show that

\begin{eqnarray*}
& N\frac{e _{0}}{kT} & \equiv  {\frac {{ 2 \pi }^{2}}{t}}  
   =  \frac{1}{( {\phi}^{2}-m)} ~ + \\  &  ~ &  {\frac {\Psi \left( \sqrt {m}-\phi +1 \right) -\Psi \left( -\sqrt {m}-\phi+1
 \right) +\Psi \left( \sqrt {m}+\phi+1 \right) -\Psi \left( -\sqrt {m}+\phi+
1 \right) }{2\sqrt {m}}}
\end{eqnarray*}
Here $\Psi(z)$ is the diGamma function \cite{GR}.
From this expression we see that the reduced chemical potential is expressible entirely in terms of the reduced angular velocity and the reduced temperature, thus confirming our choice of crossover temperature  $T_{B}\equiv Ne_{0}/(2 \pi ^{2} k)$.\\

The \textbf{reduced anomalous angular momentum}, $\ell \equiv L_{anom}/N \hbar$ can be similarly expressed:

\begin{equation}
\ell=-\sqrt {m}+ \frac{t}{2 \pi ^{2}} \left( \Psi \left( \sqrt {m}+\phi+1 \right) -\Psi
 \left( -\sqrt {m}-\phi+1 \right) - \frac{1}{ \sqrt{m}+\phi } \right)
\end{equation}

\section{Bloch Oscillations - Numerical Results}

Below we will summarize the numerical results we obtained using MAPLE. The procedure is to find the chemical potential given the number of particles N using the equation

\begin{eqnarray}
N=\sum_{n=-\infty}^\infty  \frac{1}{exp[(e_{0}(n-\phi)^2-\mu)/kT] -1}
\label{eq:N}
\end{eqnarray}

\begin{figure}[htbp]
  \begin{center}
    \epsfig{file=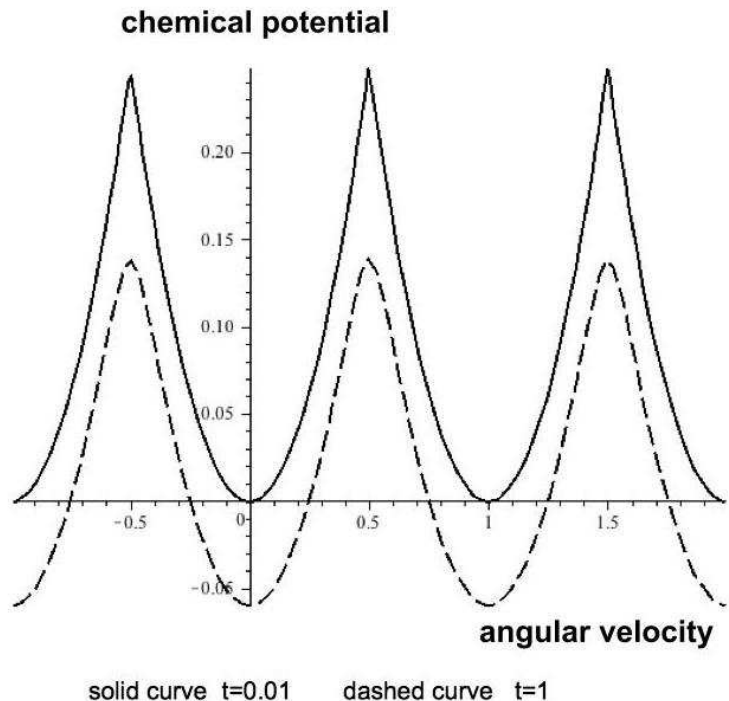, width=10cm}
    \caption{Reduced Chemical Potential vs. Reduced Angular Velocity}
    \label{fig:chem}
  \end{center}
\end{figure}

In figure (\ref{fig:chem}) we see plots of the chemical potential as a function of the reduced angular velocity - $\phi=\omega / \omega_{0}$ - for two temperatures: $T_{B}/100$ (solid curve) and $10T_{B}$ (dashed curve). The chemical potential is expressed in units of $e_{0}$. Note that the chemical potential oscillates with great amplitude in conjunction with  the Bose tendency to keep particles in the same state. This strong variation is in great contrast with the situation for Fermions, for which there is a much smaller, though sometimes important, variation that needs to be taken into account \cite{joe2} ~\cite{foot}.

\begin{figure}[h]
  \begin{center}
    \epsfig{file=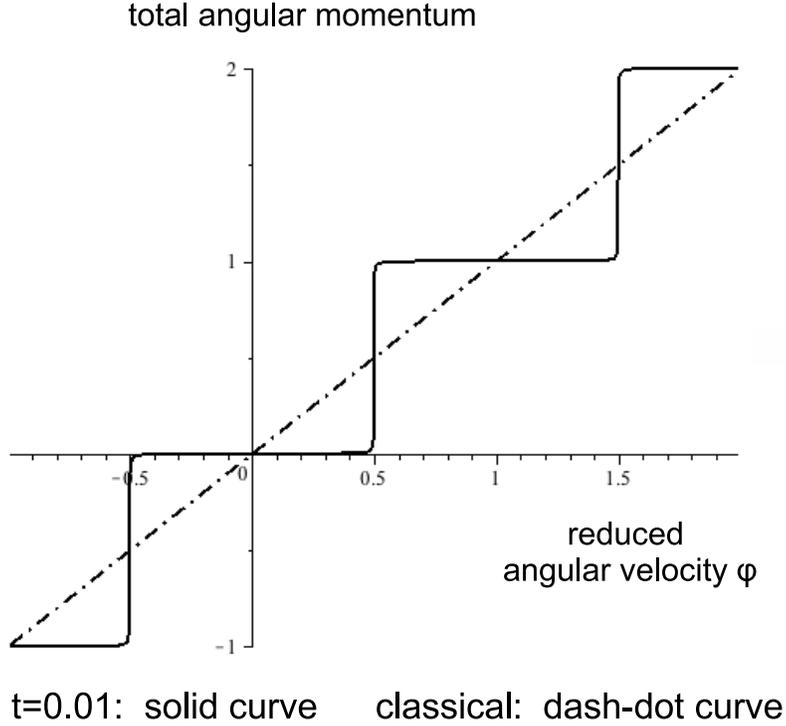, width=13cm}
    \caption{Total Angular Momentum vs. Angular Velocity}
    \label{fig:totmom}
     \end{center}
    \end{figure}

\begin{figure}[h]
  \begin{center}
    \epsfig{file=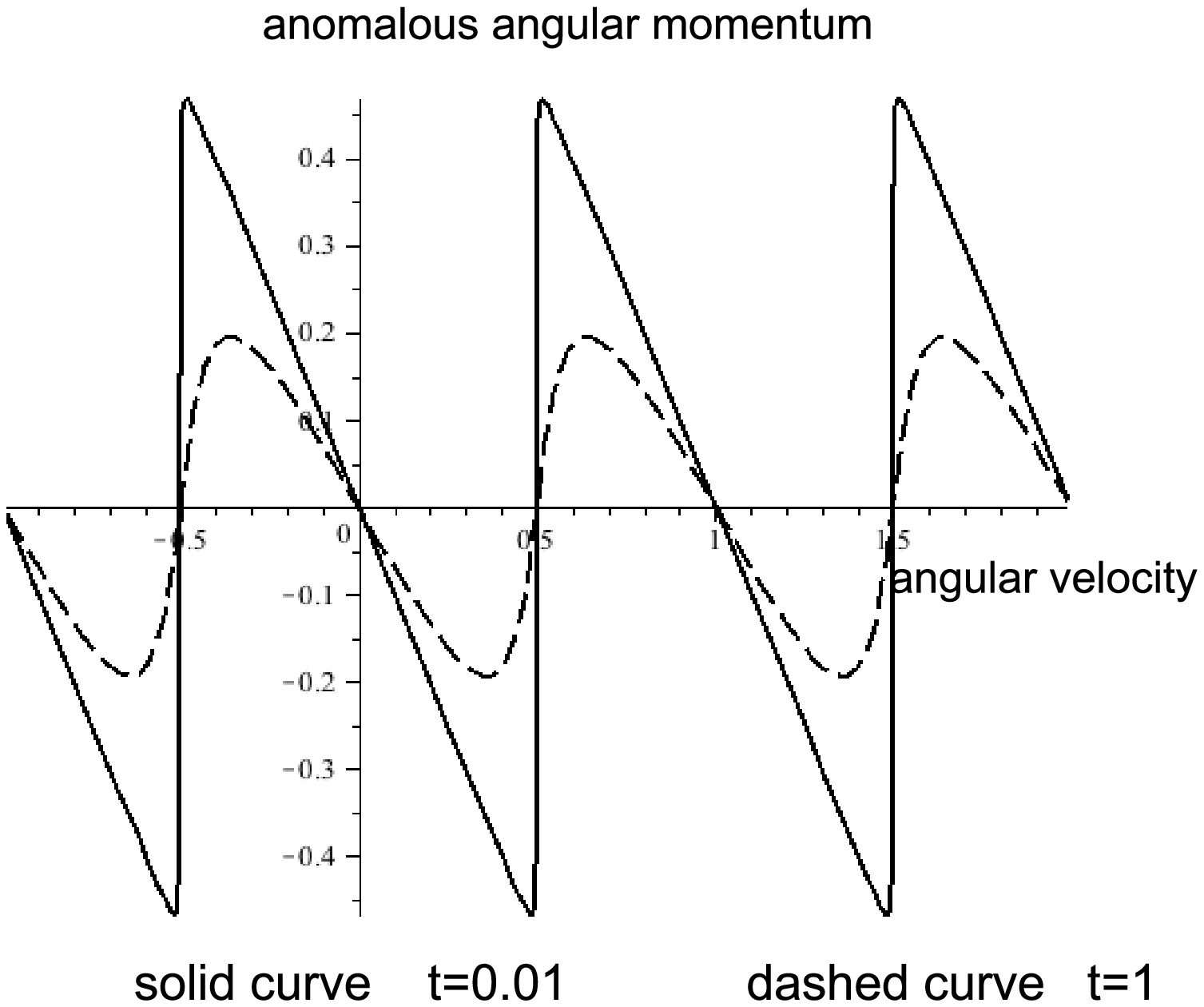, width=13cm}
    \caption{Anomalous Angular Momentum vs. Angular Velocity}
    \label{fig:mom}
     \end{center}
    \end{figure}

In figure (\ref{fig:totmom}) we plot the total angular momentum in units of $N\hbar $ as a function of the reduced angular velocity $\phi$, for the temperature $T_{B}/100$ (solid curve) and for the classical regime(dash-dot curve).  In figure (\ref{fig:mom}) we plot the anomalous angular momentum in units of $N\hbar $ as a function of the reduced angular velocity $\phi$, for two temperatures: $T_{B}/100$ (solid curve) and $10T_{B}$ (dashed curve).

\section{Josephson Oscillations}
 As shown by Bloch \cite{bloch}, Josephson oscillations in a conductor can be explained in terms of the Bloch oscillations of the current in the presence of a time dependent magnetic flux. Generally,
 \begin{eqnarray}
V= \frac{1}{c}\frac{\partial \Phi}{\partial t}
\end{eqnarray}

If the voltage is constant, the flux increases linearly in time, so that
  \begin{eqnarray}
\phi= \frac{cVt}{\Phi_{0}} = \frac{eV}{h}t
\end{eqnarray}

Then, if the electrons remain in quasi-thermodynamic equilibrium, the current will oscillate with the Josephson frequency, given by

 \begin{eqnarray}
f_{J}=\frac{eV}{\hbar}
\end{eqnarray}
\newpage
Now let us turn to the ring of an Ideal Bose gas. The corresponding experimental condition is to have a constant torque $\tau$ applied to the ring  wall. We must remember that the resulting angular velocity of the wall is not the angular velocity of the gas. However, the angular velocity of the wall determines the state of state of the gas assuming, as above, that quasi-static thermodynamic equilibrium is maintained. We have with a constant torque
 \begin{eqnarray}
L(\phi)=\tau t
\end{eqnarray}
Thus,
 \begin{eqnarray}
\phi (t)= L^{-1} (\tau t)
\end{eqnarray}
where $L^{-1}$ is the inverse function of $L$.
To obtain the corresponding Josephson frequency we use the fact that $L_{anom}(\phi)$ is a periodic function of $\phi$:

 \begin{eqnarray}
L_{anom}(\phi +1)=L_{anom}(\phi))
\end{eqnarray}

Since
 \begin{eqnarray}
L_{anom}(\phi)=L(\phi)-L_{class}(\phi)=\tau t - N\hbar \phi
\end{eqnarray}
we easily find that the Josephson frequency is
 \begin{eqnarray}
f_{J}=\frac{\tau}{N \hbar}
\end{eqnarray}

\begin{figure}[htbp]
  \begin{center}
    \epsfig{file=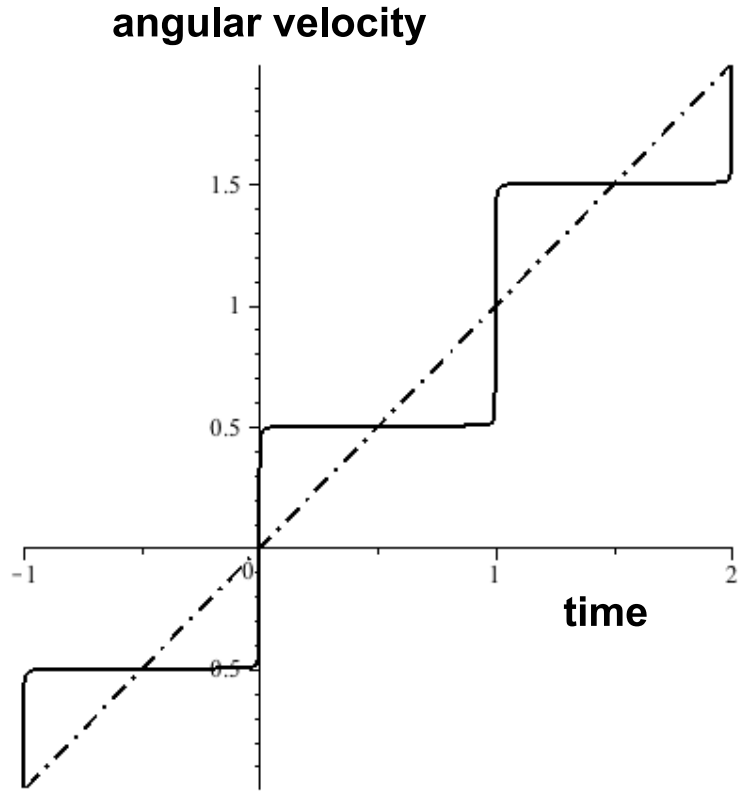, width=10cm}
    \caption{Angular Velocity vs. Time}
    \label{fig:AngvelTime}
  \end{center}
\end{figure}

A plot of the angular velocity vs. time is shown in figure (\ref{fig:AngvelTime}) for a temperature $T_{B}/100$. The time axis is in units of the Josephson period and the angular velocity is in units of the quantum of angular velocity.  The dash-dot curve is the classical result, for which the entire gas moves with the wall, so that the angular velocity is linear in time.

Let us try to make sense of the graph.  We will assume absolute zero, for which the jumps in the graph are discontinuous. First, we note that from zero to one time unit, $t_{J}=1/f_{J}$, the wall has one-half the quantum of angular velocity - that is $\omega_{_{0}} /2$ - whereas a single particle has an angular velocity that must an integral number of quanta.  Furthermore, in the course of this time interval, the total change in angular momentum is $\Delta L = \tau t_{J}=N \hbar$, which corresponds to all of the particles having a single quantum of angular velocity. Therefore, during this time interval, the particles are continuously making a transition from the $n=0$ state to the $n=1$ state. At the end of the time interval, the entire gas is moving twice as fast as the walls!  The subsequent behavior is then obvious and will not be discussed here.

\section{Proof of Bloch \& Josephson Oscillations in a Ring of Interacting Particles}
 \textbf{THEOREM}:    \emph{In the quantum regime, a ring of identical \textbf{interacting particles}, whether they be Bosons or Fermions, has an anomalous component of the angular momentum in response to a fixed angular velocity of the wall - a component that is periodic in the angular velocity, with a period $\omega_{0}$.}\\

   The proof makes use of the technique used by Byers and Yang \cite{by} in proving that in the presence of a uniform magnetic field,  magnetic flux is quantized in a hollow superconducting cylinder. \cite{periodiccurrent}
\\

 \textbf{Lemma 1}: \\
 The energy eigenvalues in the rotating frame are periodic functions of the angular velocity, with a period equal to the quantum of angular velocity.\\

\textbf{ Proof:}\\
  The coordinates of the particles can be taken to be the angles $\{\theta_{j} \}$, with $j=1...N$.  The Schroedinger equation in the lab frame is given by

     \begin{eqnarray}
\widehat{H}\Psi=E\Psi
\end{eqnarray}
  where the Hamiltonian operator $\widehat{H}$ in the inertial lab frame is given by

   \begin{eqnarray}
\widehat{H}=-\frac{\hbar^{2}}{2mR^{2}}\sum_{j=1}^N\frac{\partial^{2}}{\partial\theta_{j}^{2}} + V(\{\theta_{j}\})
\end{eqnarray}
In the rotating frame of reference, the Schroedinger equation is given by
     \begin{eqnarray}
\widehat{H}_{rot}\Psi_{rot}=E_{rot}\Psi_{rot}
\end{eqnarray}
  where the Hamiltonian operator $\widehat{H}_{rot}$ is given by

   \begin{eqnarray}
\widehat{H}_{rot}=\frac{\hbar^{2}}{2mR^{2}}\sum_{j=1}^N
(-i\frac{\partial}{\partial\theta_{j}} - \phi)^{2}
+   V(\{\theta_{j}\})
\end{eqnarray}
Note that the energy eigenvalue is a function of $\phi$: $E_{rot}=E_{rot}(\phi)$. Also, the boundary condition on the wave function is
     \begin{eqnarray}
\Psi_{rot}(\theta_{i}+2\pi)=\Psi_{rot}(\theta_{i})
\end{eqnarray}
for any $\theta_{i}$.\\
Now let us define the function $\Psi_{rot}'$:

    \begin{eqnarray}
\Psi_{rot}=e^{i\phi \sum_{j}\theta_{j}} \Psi_{rot}'
\end{eqnarray}
This new function satisfies the equation (note that the Hamiltonian corresponds to the inertial frame and is independent of $\phi$):

    \begin{eqnarray}
\widehat{H}\Psi_{rot}'=E_{rot}(\phi) \Psi_{rot}'
\end{eqnarray}
where the energy eigenvalue depends upon $\phi$ because the boundary condition on $\Psi_{rot}'$ is
     \begin{eqnarray}
\Psi_{rot}'(\theta_{i}+2\pi)= e^{-2 \pi i \phi}  \Psi_{rot}'(\theta_{i})
\end{eqnarray}
for a given $\theta_{i}$, with all other $\theta_{j}$ kept fixed.\\

We then note that if $\phi$ changes by unity $\phi \rightarrow \phi +1$, the boundary condition doesn't change. Since the boundary condition determines the specific solution to the Schroedinger equation and hence the eigenvalue, we see that the eigenvalues, with subscript label $s$, satisfy $E_{rot,s}(\phi +1) = E_{rot,s}(\phi)$.\\
 \textbf{QED}\\

\textbf{Lemma 2:}\\
The partition function in the rotating frame is a periodic function of the angular velocity, with a period equal to the quantum of angular velocity.\\

\textbf{Proof:}\\
This result follows automatically from the fact that the partition function in the rotating frame is given by
   \begin{eqnarray}
Z_{rot}=\sum_{s}e^{- E_{rot,s}/kT}
\end{eqnarray}\\
\textbf{Lemma 3:}\\
The angular momentum in the rotating frame, $L_{rot}\equiv L_{anom}$  is given by
   \begin{eqnarray}
L_{anom}= kT  \frac{\partial \ln  Z_{rot}}{\partial \omega}=-\frac{\partial F_{rot}}{\partial \omega}
\end{eqnarray}
where $F_{rot}$ is the free energy in the rotating frame.\cite{cfcurrent}\\

\textbf{Proof:}\\
The partition function in the rotating frame can be expressed as
   \begin{eqnarray}
Z_{rot}=Tr~ e^{- \widehat{H}_{rot}/kT}
\end{eqnarray}
while the Hamiltonian in the rotating frame can be expressed as
  \begin{eqnarray}
\widehat{H}_{rot}=\frac{\widehat{L}_{rot}^{2}}{2I}
+   V(\{\theta_{j}\})
\end{eqnarray}
where
  \begin{eqnarray}
\widehat{L}_{rot}= \widehat{L} - I\omega =\hbar \sum_{j} (-i\frac{\partial}{\partial \theta_{j}} ) -I\omega
\end{eqnarray}

It easily follows that

 \begin{eqnarray}
L_{anom}=\frac{1}{Z_{rot}}Tr~ e^{- \widehat{H}_{rot}/kT}\widehat{L}_{rot} = kT  \frac{\partial \ln  Z_{rot}}{\partial \omega}
\end{eqnarray}

The original theorem follows trivially since the periodicity of the partition function is passed on to the angular momentum. In addition, since the energy eigenvalues can be shown to be \emph{even} functions of the angular velocity, the angular momentum is an \emph{odd} function of the angular velocity. And finally, we should note that while we have demonstrated the periodicity of the angular velocity, there is no guarantee that its amplitude is non-vanishing.
\section{Summary and Discussion}
 We have shown that a ring of an Ideal Bose gas exhibits both Bloch and Josephson oscillations at a low temperature. Let us consider a concrete numerical example to assess the feasibility (albeit at present remote) of observing these oscillations. We will assume a spin-polarized condensate of $N=10,000$ atoms of hydrogen as a Boson with low mass -- $1.7\times 10^{-24}gm$ and a ring radius of $0.1mm$.  We obtain the following:
 \begin{center}
 $\omega_{0}=5.9$ rad/sec,~~~$e_{0}=2.9\times10^{-27}$ ergs, ~~~ $T_{B}=30~nK$.
 \end{center}
 It is difficult to imagine how one could observe this anomalous behavior with current experimental techniques. While there has been great progress in confinement of BE condensates, the anomalies presented here require the presence of a confining toroidal wall that can be controllably rotated.\\
 Josephson oscillations of an electric current in a ring are the natural expansion of the long-known oscillations (usually referred to as "Bloch oscillations") in the velocity of electrons in a periodic lattice that are driven by an electric field $\mathcal{E}$. Here, as long as electrons remain in one band (Zener tunneling being absent) the electrons move from one end of a band to a zone edge, where they are reflected to the opposite edge. The frequency is given by $f=e\mathcal{E}a/h$, where $a$ is the lattice spacing. In the case of ring, the lattice spacing of a periodic lattice is replaced by the circumference of the ring. The potential V in the Josephson frequency $f_{J}=eV/h$ is related to the electric field through $\mathcal{E}a=V$.  Similarly, the anomalous behavior presented in this paper has its counterpart in motion of atoms in a periodic lattice. In fact, T. Salger et al \cite{salger} have recently presented evidence of Bloch oscillations of a BE condensate Rb atoms moving in a periodic optical lattice. Here, the optical lattice is accelerated so as to produce an effective external force.


\begin{thebibliography}{99}
\bibitem{chempot} In deriving this result, it is assumed that the chemical potential is independent of the flux. This will be so for the Ideal Fermi gas when the temperature is much less than the Fermi temperature $\mu /k$. Note that the crossover temperature is much less than the Fermi temperature by a factor of about N.
\bibitem{foot} It is interesting that in the quest for a solution to equation (\ref{eq:N}) using MAPLE, there are multiple solutions corresponding to the various 'bands'. Thus, one must restrict the search to solutions with the lowest energy.
\bibitem{periodiccurrent} It is easy to show that the current is a periodic function of the flux, though Byers and Yang don't seem to state so explicitly.
\bibitem{cfcurrent} This result is analogous to the expression for the electric current in a  ring:\\
         $I= -\partial F/\partial \Phi$.
\end{thebibliography}
\end{document}